\documentstyle[aps,epsf,twocolumn,prb]{revtex}
\begin{document}
\draft

\twocolumn[\hsize\textwidth\columnwidth\hsize\csname @twocolumnfalse\endcsname
%%%%%%%%%%%%%%%%%%%%%%%%%%%%%%%%%%%%%%%%%%%%%%%%%%%%%%%%%%%%%%%%%%%%%%%%%%%%%%%
\title{Numerical Study of the Three-Dimensional Gauge Glass Model}

\author{J. Maucourt and D. R.  Grempel}

\address{D\'epartement de Recherche Fondamentale sur
la Mati\`ere Condens\'ee \\
SPSMS, CEA-Grenoble, 17, rue des Martyrs, 38054 Grenoble
Cedex 9, France}

\date{\today}
\maketitle
\widetext
\begin{abstract}
\noindent
%1         2         3         4         5         6         7         8
%1234567890123456789012345678901234567890123456789012345678901234567890123456780
We investigate numerically the finite-size scaling properties of the 
domain wall energies in the three-dimensional gauge glass model. 
From the analysis of results obtained for systems of linear sizes $3\le L\le 8$ 
we conclude that the stiffness exponent of the model 
is positive. This implies the existence of a stable ordered phase at low but finite temperatures.

\end{abstract}

\pacs{PACS numbers: 64.60.Cn, 75.10.Nr, 05.70.Jk}

]
\narrowtext

It has been suggested that in type-II superconductors
 at low temperatures defects may pin the flux lines 
at random positions thus destroying the Abrikosov vortex lattice. 
This leads to a  new type of superconducting state, the 
{\em vortex glass}\cite{fisher1,fisher2}, in which the phase of the 
superconducting order parameter is random in 
space but frozen in time, much in the same way magnetic moments are frozen 
in the low-temperature phase of spin glasses. The simplest system 
expected to have an ordered phase analogous to the vortex glass is 
the gauge glass model, originally introduced to describe disordered arrays 
of Josephson junctions in an external magnetic field\cite{ebner,granato}. 
This model is defined by the Hamiltonian 
\begin{equation}
H=-J\sum_{ij} \cos \left(\theta_i-\theta_j-A_{ij}\right),
\label{hamil}
\end{equation}
where $\theta_i$ is the phase of the order parameter at the $i$-th 
site of a simple cubic lattice and the sum 
runs over all pairs of neighboring sites. The energy scale is set by the 
coupling constant $J$ and the lattice spacing is 
identified with the typical distance between vortices\cite{fisher3}. The phase 
shifts $A_{ij}=(2\pi/\Phi_{0})\int_{i}^{j}\vec{A}\cdot d\vec{l}$ 
where $\vec{A}$ is the vector potential of the applied magnetic field and 
$\Phi_{0}$ the flux quantum. The effects of the  
 disorder in the positions of 
the vortices are incorporated by 
taking the phase shifts as independent quenched random variables. The situation 
that interests us, where the disorder and the external field are large, 
may be modeled by taking $A_{ij}$ uniformly distributed in the interval 
$[0,2\pi]$. The three-dimensional gauge glass model has been 
extensively studied 
numerically by Monte Carlo simulation \cite{huse,reger,wengel} and 
finite-size scaling of defect wall energies \cite{reger,gingras,kost} 
the important issue being whether a thermodynamically ordered phase can exist
 at finite temperature in this system. Although the results of the 
earlier Monte-Carlo studies\cite{huse,reger} of the model were consistent
 with the existence of a low-temperature vortex glass phase, they could not 
rule out a zero-temperature transition  
since only small systems could be brought to equilibrium below $T\sim 0.6 J$. 
Stronger evidence in favor of a finite-temperature
 transition has been obtained in recent simulations based on the 
vortex representation of the problem\cite{wengel} in which it was 
found that the transition 
temperature may be as high as $T_{\rm c}=0.93 J$.  
On the other hand, the domain-wall renormalization-group (DWRG) studies 
performed so far \cite{reger,gingras,kost} were inconclusive, the sizes of 
the systems studied being too small and the 
statistical error too
 large to decide unambiguously whether the lower critical dimension 
of the model is above or below three.
In this paper we reexamine this problem by means of a 
 DWRG study of model (\ref{hamil}) with an algorithm that we 
have recently proposed and applied to the $XY$ spin-glass model in three 
dimensions\cite{us}. 
This algorithm allows us to  
study lattices substantially bigger  than with conventional methods as 
well as to improve upon the statistics. 
In the defect wall
 method\cite{banavar,mac} the energy cost $W$ of introducing a 
domain wall in the system is studied as a
function its linear size $L$. In the scaling regime one 
finds \cite{banavar,mac} $W\sim L^{\theta}$ where the 
stiffness exponent $\theta$ may be positive or negative depending on 
whether the system is above or below its lower critical dimension $d_{\rm c}$. 
From the results obtained for our five largest sizes ($4\le L \le 8$) 
we find the value $\theta_{\rm GG}=0.077 
\pm 0.011$ for the gauge glass model.  This result implies 
that a stable ordered phase exists at low but finite temperatures.

To determine the domain-wall energy one computes the differences 
$\Delta E=E_{\rm P}-E_{\rm A}$ between the ground-state energies 
corresponding to  periodic (P) and anti-periodic (AP) boundary conditions 
along some direction  for an 
ensemble of systems of size $N_{\rm s}=L^3$. The boundary conditions along
 the two remaining directions are kept fixed. For 
sufficiently large systems the
 distribution of energy differences differences ${\cal P}(\Delta E,L)$ is
  expected 
to have the scaling form\cite{mac}
\begin{equation}
\label{distribution}
{\cal P}(\Delta E,L)=L^{-\theta}\tilde{{\cal P}}(\Delta E L^{-\theta}).
\end{equation}
The width of the distribution, 
$W(L)=\langle\Delta E^2\rangle_{A_{ij}}^{1/2}\sim L^{\theta}$, is interpreted as
 the effective 
coupling constant between blocks of $N_{\rm s}$ 
sites\cite{banavar,mac,morris}.
If $\theta>0$, the rigidity of a block diverges with its size, 
which indicates that the system has long-range order. If the
 stiffness exponent is negative, the correlation length diverges at 
$T=0$ with $\xi\sim T^{-\nu}$ and $\nu=1/|\theta|$ \cite{morris}. 

The ground state of the gauge glass model is given by 
the absolute minimum of (\ref{hamil}) 
subject to the appropriate boundary conditions. In the presence of 
disorder the extremal conditions $\partial H/\partial\theta_i=0\  \forall i$ 
have in general a very large 
number of solutions whose presence greatly complicates the task 
of searching that with the lowest energy. In the spin-quench 
algorithm\cite{walker}(SQA) usually employed to solve this type of 
problem,
 long sequences of metastable states are randomly 
generated among which one will find the ground-state provided   
 the number of trials is sufficiently large. Since the number of metastable 
 states of a frustrated 
system increases exponentially with its size\cite{morris}, so does  
the number of trials required. This limits 
the maximum size of the systems that can be studied using 
this method in practice.
We have recently proposed a far more efficient algorithm for the search 
of ground states\cite{us}. It is based upon the  
 morphological characteristics of the 
low-lying states of frustrated $XY$ models as revealed by detailed 
examination of numerous examples\cite{gawiec}. 
For a given realization of the disorder in Eq.\ref{hamil}, 
the low-energy configurations are characterized by the existence of 
regions where the order parameter varies smoothly (domains), and others 
where the spatial distribution of the phases looks pretty much random. 
The former exist in parts of the sample where frustration is 
low, the latter where it is high. As it turns out\cite{gawiec},
 the position, size and 
shape of the domains are mostly determined by the  
realization of the disorder and are essentially the same 
 for all the low-energy states. Aside from smooth 
distortions of the order parameter, 
the essential differences
 between any two such states are almost  
rigid rotations of the individual domains, accompanied by large amplitude 
rearrangements of the 
phases in the frustrated regions between them. Stationary states in 
which the domain structure is disrupted do exist, but their energy 
is much higher.
In our method, sequences of low-energy configurations are generated 
recursively in such a way that the domain structure is preserved 
at each step. The result is 
a reduction of the probability of appearance of high-energy 
configurations in the sequence and a corresponding enhancement of that of 
finding the ground-state or states lying nearby in energy.
The procedure is as follows \cite{us}.  
The first state in the sequence, $\{\theta^{(0)}\}$, 
is obtained by a conjugate-gradient minimization (CGM) of the energy  
(\ref{hamil}) starting from a random distribution of phases. 
New states are generated by iterating the following steps.
i) Sites are divided 
in two classes  
according to whether the `local field' $h_{i}=-J\sum_j 
\cos\left(\theta_i-\theta_j-A_{ij}\right)$ in the $n$-th 
configuration $\theta^{(n)}$is
greater or smaller than a threshold value $h_{\rm t}$ chosen as 
explained  
below. The sites in the first group constitute the domains. 
ii) Correlations between the domains and the rest of the 
system are destroyed by a random rigid   
rotation of the former.
iii) A fraction $p$ of the $N_{\rm w}$ sites in weak local 
fields are picked at random and their phases reset to arbitrary values. 
iv) The energy of the subsystem formed by the domains is minimized 
{\em with the phases on the remaining sites fixed}. 
v) The state resulting from the previous step is allowed to relax 
by performing a CGM of the {\em total} energy of the system. The 
outcome is the next state in the sequence, $\{\theta_{\rm n+1}\}$.
vi) The energy of this state is stored and, eventually, $h_{\rm t}$ is 
rescaled.

The efficiency of this algorithm depends upon the chosice of 
the parameters $h_{\rm t}$ and p. The threshold field fixes  
the degree of homogeneity required 
of a region for it to be classified as a domain. 
If it is too high or if $p$ is too large, too  
many sites are involved in step iii) and the domain structure 
is disrupted just as in the SQA 
where all the phases are randomly reset at each step. 

\begin{figure}
\epsfxsize=3.5in
\epsffile{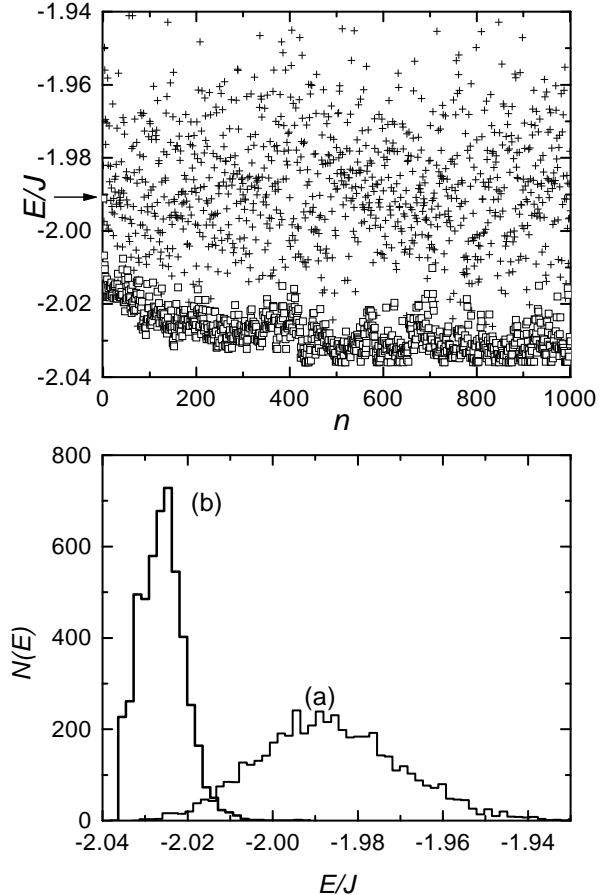}
\caption{Upper panel : Energies of some of the metastable states 
of a realization of the 
gauge glass model on a 512-site 
3D lattice. Results are plotted as a function of 
iteration number for the spin quench algorithm (crosses) and for   
our algorithm (squares). The arrow indicates the energy 
of the first state in our sequence.
Lower panel :  Distribution of the energies of two series 
of 5 000 metastable states each obtained with the SQA (a) and with
our algorithm (b).}
\label{fig1}
\end{figure}

If, on the contrary, $h_{\rm t}$ or $p$ are too small, the algorithm  
gets trapped in phase space and all the states in the sequence are close to the
 initial one. The two parameters must therefore 
 be continuously readjusted in the course of the simulation to ensure    
good performances.
We have empirically found that the algorithm performs at its best for 
large samples 
when the number of sites involved in step iii) $p N_{\rm w} \equiv 
n\sim 0.05 N_{\rm s}$. If at some stage of the iteration $N_{\rm w}<n$
 we consider that the threshold field is too low and too many sites 
are being included in the domains. We then rescale it upwards, 
$h_{\rm t} \to (1+\alpha)\  h_{\rm t}$ with $\alpha \sim 0.05$, and we randomly 
reset the phases on all the sites where the local field is weak. 
If $n\le N_{\rm w}\le 2n$,  
 $h_{\rm t}$ is unchanged and the phases are updated on just
 $n$ randomly chosen sites. Finally, if $N_{\rm w}>2n$ we consider that the
  threshold field is too high and we rescale it downwards according 
to $h_{\rm t} \to (1-\alpha) \ h_{\rm t}$.
We find that, in practice, the field stabilizes itself after a few iterations
 and oscillates about a value that, in the case of the simulations 
 reported here, is $\sim 2.6 J$.

In the upper panel of Fig.\ref{fig1} we show the energies 
of two series of 1000 minima  of (\ref{hamil}) for a particular realization  
of the disorder. Data were obtained for a  
 3D lattice of $512$ sites using periodic boundary conditions.
The crosses represent states obtained with the SQA and the squares 
are the outcome of the first thousand iterations of our algorithm. 
The arrow points at the energy 
of the first configuration of our sequence. It can be seen that, whereas 
the conventional algorithm randomly samples 
the whole of phase space, 
our method seems to mostly explore the deepest valleys. 
Notice that during the first
five hundred or so iterations the typical energy of the states in 
the sequence
decreases continuously after which it stabilizes  
in a region of energies that is hardly ever visited by the SQA. It is 
important to check that the configurations that enter in the sequence come 
from well 
separated regions of phase space rather than from a particular valley 
where the algorithm would be trapped. This may be done simply
 by monitoring the evolution 
of the overlap of the successive configurations with a particular 
one that is
chosen as reference. 
The lower panel of Fig.\ref{fig1} shows histograms obtained 
after five thousand iterations of the two algorithms. It can be seen 
that the histogram obtained with our method is much narrower and
 centered at a much lower
energy. The overall features of the distribution of energies 
shown in  Fig.\ref{fig1} are quite similar to those 
recently found for the $\pm J$ spin glass model\cite{us}.
It is remarkable that about twenty percent of 
the states 
found using our algorithm in this example have never been 
generated by the SQA. 
Our lowest 
energy state, at $E=-2.036 J$, appears $\sim 200$ times 
in the sequence. The configurations of the states that have this energy 
are related to each other 
by uniform rotations. In between them, the algorithm generates states 
that are in far away regions of phase space.
We believe that this state is the ground state of this particular realization.

\begin{figure}
\epsfxsize=3.5in
\epsffile{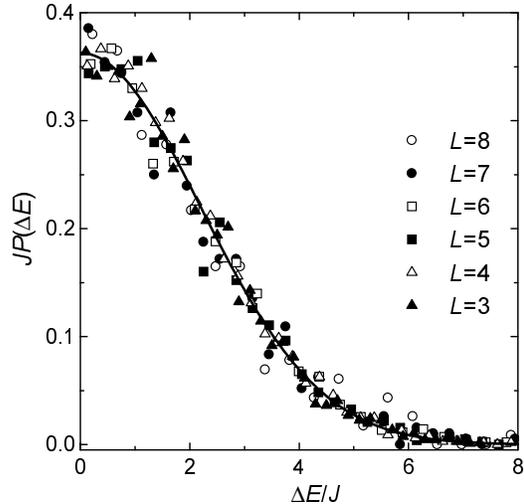}
\caption{Probability distributions of the differences of the ground-state 
energies with periodic and antiperiodic boundary conditions. 
The solid line is the result of a
 gaussian fit of the ensemble of the data.}
\label{fig2}
\end{figure}

In order to study 
the scaling properties of defects energies in the gauge glass 
model we have applied the above method to compute 
 ground-state energies with periodic and antiperiodic boundary conditions 
 for systems of $L^3$ sites with $3\le L \le 8$.
Only systems with $L\le 5$ had been investigated 
previously\cite{gingras,kost} .
We generated sequences of states containing 500 ($L$=3), 
800 ($L$=4), 1000 ($L$=5), 2000 ($L$=6), 3000 ($L$=7) and 5000 ($L$=8) 
elements, respectively. Disorder averages were taken over 25600 
($L$=3), 6400 ($L$=4), 
2560 ($L$=5,6), 640 ($L$=7) and 256 ($L$=8) samples, respectively.
The normalized distributions of the differences 
$\Delta E=|E_{\rm P}-E_{\rm AP}|$ obtained numerically for the 
different sizes are shown in Fig.\ref{fig2}. Detailed examination of the 
results shows that the differences between the curves for different sizes
 are of the same order of magnitude as the
statistical error bars. Because of this we were not able to determine 
the stiffness exponent of the model by performing a 
scaling plot as Eq.\ref{distribution} suggests. 

\begin{figure}
\epsfxsize=3.5in
\epsffile{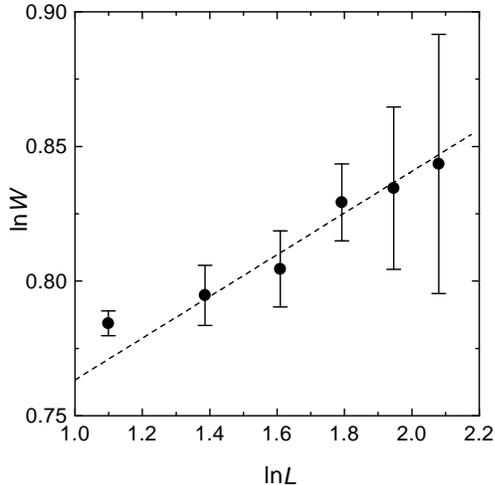}
\caption{The $L$-dependence of the  domain-wall energies for the 
gauge glass  
model on $L\times 
L\times L$ simple cubic lattices. The 
dashed line is the power-law fit discussed in the text.}
\label{fig3}
\end{figure}

The solid curve in the Fig.\ref{fig2} is a 
gaussian of width  $W=2.25 J$. The fact that we can quite 
reasonably describe the ensemble of the data using a single  
size-independent distribution is an indication 
that the lower critical dimension of the gauge glass problem is very close to 
three, as found by other authors\cite{reger,gingras,kost}. 
The $L$-dependence of the effective coupling 
$W(L)=\langle\Delta E^2\rangle_{A_{ij}}^{1/2}$ is
 shown in the log-log plot of Fig.\ref{fig3}.
As the figure shows, statistics for the 
two largest systems is still unsatisfactory but very hard to improve upon 
because 
of CPU-time limitations. Nevertheless, we can still conclude from the 
available data that 
the domain-wall energy increases slowly with length scale. 
Leaving out the point for $L=3$ which is likely to be too small a size for 
scaling to hold, we can make a power-law fit of the results. The stiffness 
exponent thus determined is $\theta_{\rm GG}=0.077 \pm 0.011$. We have 
checked this 
result by repeating the calculation for the larger sizes starting 
from different random initial configurations. The differences 
found between the 
results thus obtained fall within the statistical error bar. 
Our value for the stiffness exponent is consistent with those 
reported by Gingras\cite{gingras}($\theta_{\rm GG}=0.04\pm 0.06$) and by other 
authors\cite{reger,kost} who find $\theta_{\rm GG} \approx 0$ within their 
statistics. Ours is, to our knowledge, the first calculation in which
 the possibility $\theta_{\rm GG}\le 0$ is outside the range covered 
 by the error bar.

The results of this paper thus confort the idea that the lower 
critical dimension of the gauge glass model is slightly below three. 
The implication is that the system has a finite-temperature transition 
to an ordered state in agreement with 
the findings of the Monte Carlo 
studies of the model\cite{huse,reger,wengel}. It is interesting to 
notice that whereas the smallness of $\theta_{\rm GG}$ would lead one 
to naively expect a very low transition temperature, the Monte 
Carlo data indicate that $T_{\rm c}\sim {\cal O}(J)$. This is a 
somewhat puzzling result that deserves further investigation.  

The calculations presented here have been 
done on a 256-processor CRAY T3E parallel computer at the 
`Centre Grenoblois de Calcul Vectoriel'. We thank the staff for their 
technical help.


\begin{references}

\bibitem{fisher1}
M. P. A. Fisher, Phys. Rev. Lett. {\bf 62}, 1415 (1989).
\bibitem{fisher2}
D. S. Fisher, M. P. A. Fisher and D. A. Huse, Phys. Rev. B {\bf 43}, 
130 (1991).
\bibitem{ebner}
C. Ebner and D. Stroud, Phys. Rev. B {\bf 31}, 165 (1985).

\bibitem{granato}
E. Granato and J. M. Kosterlitz, Phys. Rev. B {\bf 33}, 6533 (1986);
Phys. Rev. Lett. {\bf 62}, 823 (1989).

\bibitem{fisher3}
M. P. A. Fisher, T. A. Tokuyasu, A. P. Young, Phys. Rev. Lett.
{\bf 66}, 2931 (1991).

\bibitem{huse}
D. A. Huse and H. S. Seung, Phys. Rev. B {\bf 42}, 
1059 (1990).

\bibitem{reger}
J. D. Reger, T. A. Tokuyasu, A. P. Young and M. P. A. Fisher,
Phys. Rev. B {\bf 44}, 7147 (1991).

\bibitem{wengel}
C. Wengel and A. P. Young, Phys. Rev. B {\bf 56}, 5918 (1997).

\bibitem{gingras}
M. J. P. Gingras, Phys. Rev. B {\bf 45}, 7547 (1992).

\bibitem{kost}
J. M. Kosterlitz and M. Simkin, Phys. Rev. Lett. {\bf 79}, 1098 (1997).

\bibitem{us}
J. Maucourt and D. R. Grempel,  Phys. Rev. Lett. {\bf 80}, 770 (1998).

\bibitem{banavar}
J. R. Banavar and M. Cieplak, Phys. Rev. Lett. {\bf 48}, 
832 (1982); M. Cieplak and J. R. Banavar, Phys. Rev. B {\bf 29}, 469 
(1984).

\bibitem{mac}
W. L. McMillan, Phys. Rev. B {\bf 31}, 342 (1985).

\bibitem{morris}
B. M. Morris, S. G. Colborne, M. A. Moore, A. J. Bray and J. 
Canisius, J. Phys. C  {\bf 19}, 1157 (1986).


\bibitem{walker}
L. R. Walker and R. E. Walstedt, Phys. Rev. B {\bf 22}, 3816 (1980).

\bibitem{gawiec}
P. Gawiec and D. R. Grempel, Phys. Rev. B {\bf 44}, 2613 (1991).

\end{references}
\end{document}